\documentclass[onecolumn,showpacs,preprintnumbers,amsmath,amssymb]{revtex4}

\usepackage{graphicx}
\usepackage{dcolumn}
\usepackage{bm}

\begin{document}

\title{Clustering Evolutionary Stock Market Model}
\author{Jie Wang}
\author{Chun-Xia Yang}
\author{Pei-Ling Zhou}
\affiliation{
Department of Electronic Science and Technology,\\
 University of Science and Technology of China,\\
 Hefei, Anhui, 230026, PR China}
\author{Ying-Di Jin}
\author{Tao Zhou}
\author{Bing-Hong Wang}
\email{bhwang@ustc.edu.cn,Fax:+86-551-3603574}
\affiliation{
Department of Modern Physics,\\
 University of Science and Technology of China,\\
 Hefei, Anhui, 230026, PR China }

\date{\today}

\begin{abstract}

As a typical representation of complex networks studied relatively
thoroughly, financial market presents some special details, such
as its nonconservation and opinions spreading. In this model,
agents congregate to form some clusters, which may grow or
collapse with the evolution of the system. To mimic an open
market, we allow some ones participate in or exit the market
suggesting that the number of the agents would fluctuate.
Simulation results show that the large events are frequent in the
fluctuations of the stock price generated by the artificial stock
market when compared with a normal process and the price return
distribution is a \emph{l\'{e}vy} distribution in the central part
followed by an approximately exponential truncation.
\end{abstract}


\maketitle

\section{Introduction}
As the research of the complex systems is getting deeper and
deeper, to find the universal rules and principles of these
systems and to answer the origination of ``complexity" become more
and more attractive \cite{53,54,55,56,57,58}. Thus, the vision of
physicists is no longer restricted in the traditional areas but
concentrates on the more comprehensive domains, leading to the
birth of many burgeoning disciplines through the interaction and
amalgamation of physics and other fields such as biology, finance,
sociology, and so on. Over the last decade, there has been
significant interest in applying physical methods in social and
economical science \cite{37}. In particular, the study of
financial market prices has been found to exhibit some universal
properties similar to those observed in physical systems with a
large number of interacting units, and some models
\cite{41,42,43,44,45} (as we know, the first stock market
simulation was performed by economist Stigler in 1964 \cite{46})
have been introduced to the financial and more recently to the
physical communities which attempt to capture the complex
behaviors of stock market prices and market agents \cite{47}.
These models, including behavior-mind model \cite{16,17},
dynamic-games model \cite{32},
  multi-agent model \cite{18,19,20,21,22,3} and so on, are based on the statistical properties \cite{23}
   of price fluctuations which should be recovered by more suitable microscopic models \cite{24}:
  (1)sharp peak and fat-tail distribution for the price changes (the return histogram) \cite{25};
  (2)the distribution of returns decays with power law in
     the tails \cite{28}, with exponent near 3; (3)price fluctuations are
     not invariant against time reversal, i.e. they show a forward-backward
     asymmetry \cite{30}.

   Among the more sophisticated approaches are dynamic
   multi-agent models
   based on the interactions of two distinct agent
   populations (``noisy" and ``fundamental" traders), which could simulate the
   price forming processes and reproduce some of the stylized observations of
   real markets, but fail to account for the origin of the universal
   characteristics. An alternative approach, the herd behavior \cite{48,49} explored in this
   paper, may be capable to induce
   the power-law asymptotic behavior in the tail of price return
   distribution with an exponent well fitting the truncated L\'{e}vy regime \cite{25}
   as found in real data \cite{44,28}. This approach has been
   formalized by Cont and Bouchaud \cite{31} as a static
   percolation model. Subsequently, this percolation model has been
   bettered by introducing a feedback mechanism between the price return $Z$ and trader activity
   $a$: $a \rightarrow a+\alpha Z$, where $\alpha$ is the factor
   denoting the sensitivity to price fluctuations. Then the
   volatility clustering can be reproduced, and all of the statistical properties of fluctuations for prices
   mentioned above could be observed.

   \par In the Cont-Bonchaud model, random percolation clusters are used as groups
   of traders. In the simple version, at each iteration each cluster buys with probability
   $a$, sells with probability $a$ or sleeps with probability $1-2a$. When
   the activity $a$ is small, there are only few clusters trading at a time
   most of the times. Therefore, the distribution $P_{b}(Z)$ of relative price
   fluctuations or ``returns" $Z$ scales as the
   well-known \cite{38,39,40} cluster size distribution of percolation theory. But when we increase the value of $a$
   , more and more clusters would make contribution to the relative
   price fluctuations and the central limit theorem suggests that
   the distribution $P_{b}(Z)$ convert from power-law tails to a
   more Gaussian shape for large systems. Price changes in the
   logarithm (return) are proportional to the difference between
   the
   supply and demand. On average, price rises or falls with equal probability and without
     correlations between consecutive steps. An assumption made in
     this model which should not be ignored is that the
     probability $a$ (activity) is set to be the same for all
     groups and remains constant through out the whole process, which may be a good strategy for simplifying a
     physical model but may be not a good regulation for
     establishing a model which we expect to reflect the various
     phenomena found in the real stock market as genuine as possible
     so it could be more helpful for us to capture the complex
     properties
     of the real world. Although in the successional studies of the
     percolation model different mechanisms are used to
     establish a self-organized model where the investor groups
     with various trading activities and sizes are formed
     automatically, few of these models considered the fluctuations of
     the traders in the market, fact that there are always agents
     who
     take part in or exit from the market due to various reasons which
     might have a serious influence on the price. Here we introduce
     a self-organized model where the activities and sizes of different investor groups are
   driven by the confliction and harmonization of the strategies
   adopted by different groups. The simulation results which well agree with
    the observations of real markets are also shown in the third section.

 \begin{figure}
\scalebox{0.37}[0.42]{\includegraphics{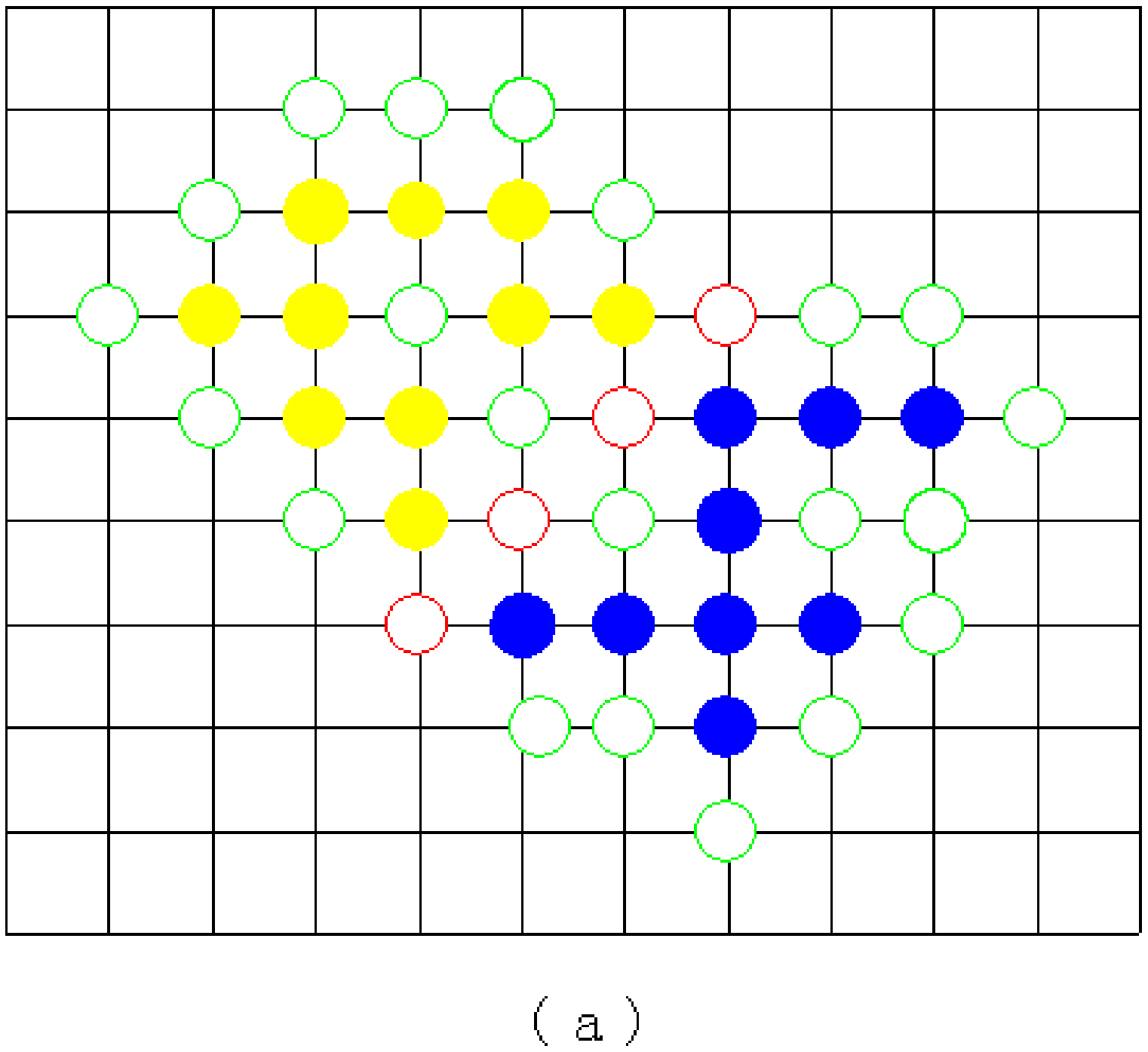}}
\scalebox{0.425}[0.425]{\includegraphics{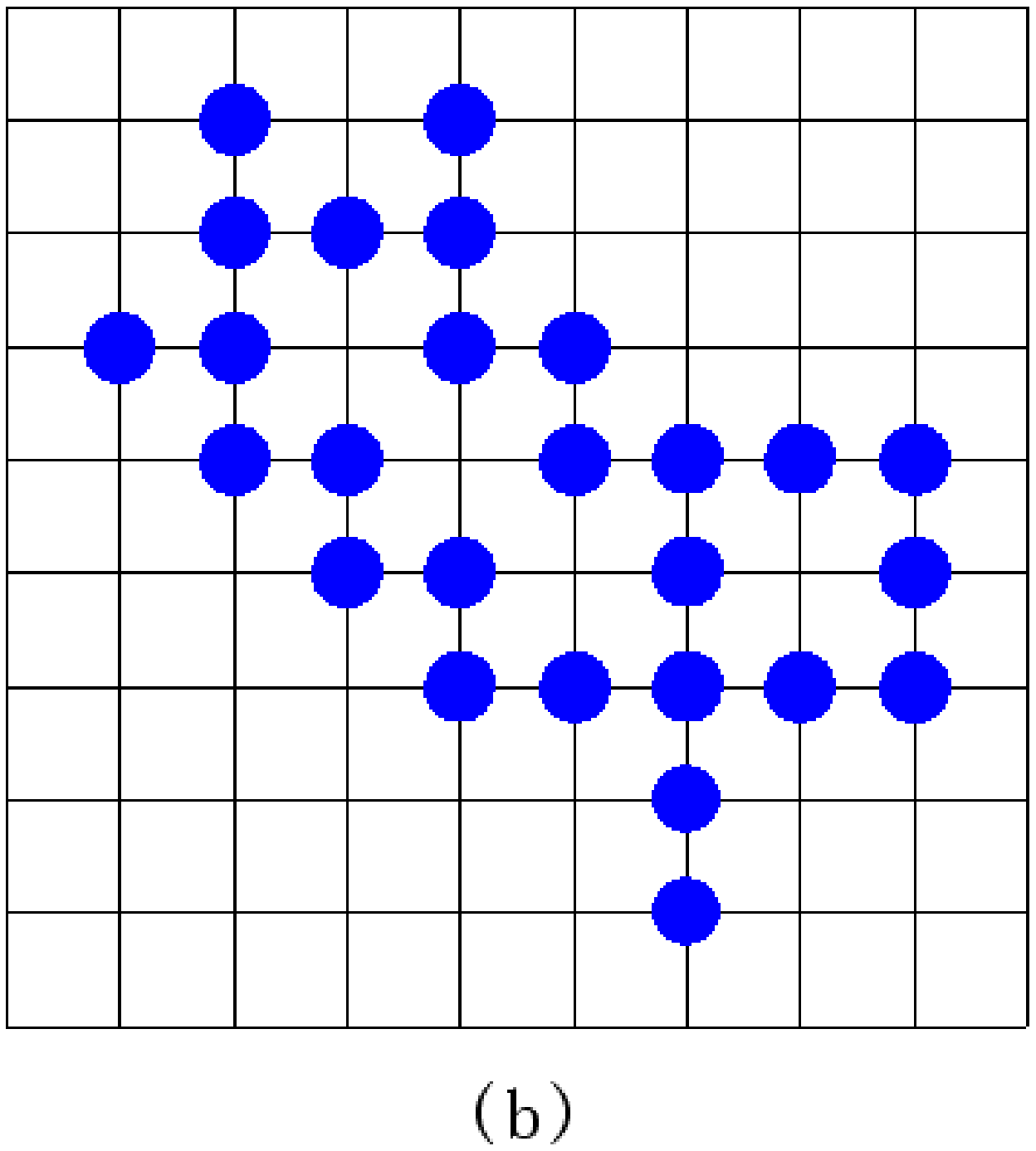}} \caption{(a)To
illustrate, this is a small-scale matrix comparing to our model
which can help us to explain what is a cluster called in this
paper. A cluster, here we call it $M$, is defined as an
aggregation of nodes that share the same information and hold the
same opinion. Moreover, in the topology, every two nodes in $M$
can reach each other. Obviously, there are two different clusters
which are represented by different colors, yellow and blue in
figure 1(a). In the process of growth, each cluster has distinct
probability to absorb new members who could occupy the frontier
and empty nodes with different probabilities determined by the
cluster which they would participate in. And the sites which the
new agents could occupy are denoted by the hollow nodes. But the
colors of the hollow nodes are distinguished which implies that
there must be some difference between the positions marked by
different colors. In fact, the positions denoted by red hollow
nodes mean that if a new member comes into the market and takes up
one of these positions, he would be puzzled for there are two
different attitudes. In other words, he comes across two different
information resources which would lead to completely different
consequences. And he operates just like a information bridge which
would stimulate the communication of clusters holding different
attitudes and even lead to a serious confliction that could
promote some certain clusters to merge others and expand to a
giant one such as the cluster shown in figure 1(b). The opposite
process called "collapse of the cluster" in this model, which
means that the nodes belonging to a same cluster are removed from
the lattice, reflects another phenomenon of the real market that
there are always some traders quitting the market due to different
reasons which may lead to a serious influence on the price.}
\end{figure}

   \section{The new model}
    \par Considering an open market which absorbs and removes traders with
   probabilities respectively in the process of trading, we draw the growth and
   collapse of clusters
   into our model based on Cont-Bouchaud model's random percolation. After every $N$ iterations of trading, each
   cluster which is defined in figure 1
   grows around itself (figure 1(a)) with
   probability $P_{d}$, collapses and annihilates with probability $P_{n}$,
   or sleeps with
   probability $1-P_{d}-P_{n}$. Once a cluster collapses, some new clusters will come
   into the market at the positions where the collapse occurred (figure 1(b)), with a fixed
   probability $P_{h}$.
We build our model as follows based on the thoughts above:

\par (1) Initiation:
 a $L\times L$ lattice is occupied randomly with probability $P_{in}$, and each cluster
 is randomly given a state: buying, selling, or sleeping, which are
 represented by 1, -1 and 0 respectively.

\par (2) Trading:
at time $t$, each agent in the market sells or buys a unit of
stock. Then we calculate the difference between the supply and
demand
\begin{equation}
r=\sum_{i=1}^{m}s_{i}
\end{equation}
where $m$ is the total number of agents who are presented in the
market at $t$ and $s_{i}$ represents the state of the $i$-th
agent. The evolution of price follows the rule:
\begin{equation}
P(t+1)=P(t)e^{r/\lambda}
\end{equation}
 where
$\lambda=N_{p}+N_{n}$. Here $N_{p}$ and $N_{n}$ denote the number
of buyers and sellers, respectively. Figure 2 shows the price time
series, which is rather similar to that of real stock market. And
the active probability $2a$ with which the agents choose buying or
selling rather than sleeping evolves following the Equ.(3):
\begin{equation}
a(t)=a(t-1)+lr
\end{equation}
where $l$ represents the sensitivity of activity $a(t)$ to the
difference between the demand and supply. And then, each cluster
buys, sells, or sleeps with probabilities $2a(t)p_{b}$,
 $2a(t)p_{s}$, $1-2a(t)$ respectively, where
\begin{equation}
p_{b}=\left\{
    \begin{array}{cc}
        \mu+\nu_{1}r, &\mbox{$r<0$}\\
        \mu+\nu_{2}r, &\mbox{$r>0$}
    \end{array}
    \right.
\end{equation}
and
\begin{equation}
p_{s}=1-p_{b}
\end{equation}
 The first term on the right hand of both of Equ.(4) denotes
 the probability with which the active one would buy rather than sell without considering
 the feedback of the price fluctuations.
 The difference between the case $r<0$ and $r>0$ is the
 coefficient of the last term on the right hand and takes into account that agents are risk adverse and
thus more impressed by a downturn than by an upturn of the market
so that the parameters $\nu_{1}$ and $\nu_{2}$ denote the
sensitivity of the agent's mentality to the price fluctuations.
(The price would fall if $r<0$ and would rise if $r>0$ according
to Equ.(2) \cite{4,5,6,7,8,52}). So the value we adopt of
$\nu_{1}$ is smaller than that of $\nu_{2}$ \cite{1}.
\par (3) Growth:
after each $N$ iterations, there are three types of evolution with
different probabilities respectively depicted in the following
segments: growing, collapsing and sleeping.
\par The first situation is that new
traders come into the market occupying the empty sites around the
old clusters, for example, cluster \textbf{G}, just as the sites
marked by hollow nodes proposed in figure 1(a) with the
probability
 \begin{equation}
 P_{d}^{g}(t+1)=P_{d}^{g}(t)+k(N_{T}-c^{g}(t))
 \end{equation}
 where $k$ is a kinetic coefficient and $N_{T}$ is a threshold parameter
 \cite{2}, and
\begin{equation}
c^{g}(t)=\sum_{j=1}^{m_{g}}(|s_{j}^{g}|)
\end{equation}
in which $m_{g}$ represents the scale of the cluster $\textbf{G}$,
in other words, the number of nodes which belong to $\textbf{G}$.
$s_{j}^{g}$ denotes the state of the $j$-th node belonging to
$\textbf{G}$.
 The probability $P_{d}^{g}(t+1)$ is obviously limited to the range
 $[0,1]$ so that we have to impose $P_{d}^{g}(t+1)=0$ and $=1$
 if the recurrence relationship Equ.(6) gives values for
 $P_{d}^{g}(t+1)<0$ or $>1$.
 If  a few of different clusters whose states are different encounter, one ( noted by $\textbf{V}$ ) will defeat others ( the total
 number is $n$, including $\textbf{V}$ ) with the probability
\begin{equation}
 P_{v}(t)=|c^{v}(t)|/\sum_{i=1}^{n}|c^{i}(t)|
 \end{equation}
 And the evolution due to Equ.(8) would lead to the consequence that the
 defeated clusters would accept the opinion and adopt the same strategy of the winner. In other words, they are annexed by the winner.
 By contrary, when the states of the encountered
clusters are all the same, they would combine and operate as a
whole.
 \begin{figure}
\scalebox{0.888}[0.888]{\includegraphics{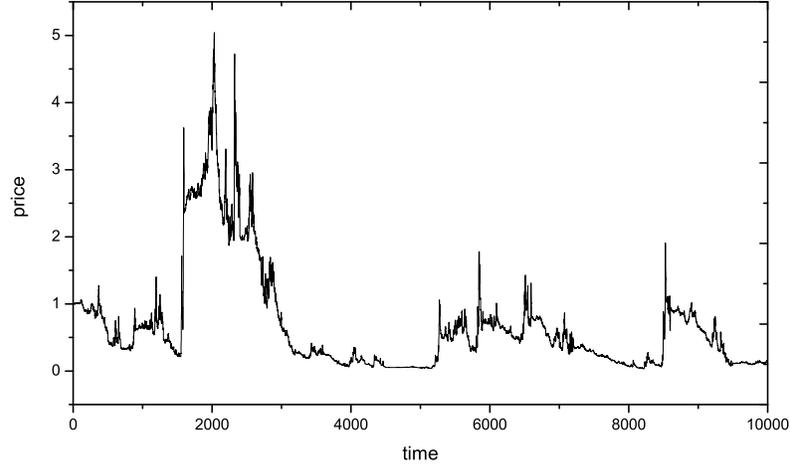}}
\caption{\label{fig:epsart}Time series of the typical evolution of
the stock price, where $P_{h}$=0.01. One can see that the trend
and fluctuations of the stock price generated by our model are
rather similar to that of real stock market.}
\end{figure}
\begin{figure}
\scalebox{1.2}[0.9]{\includegraphics{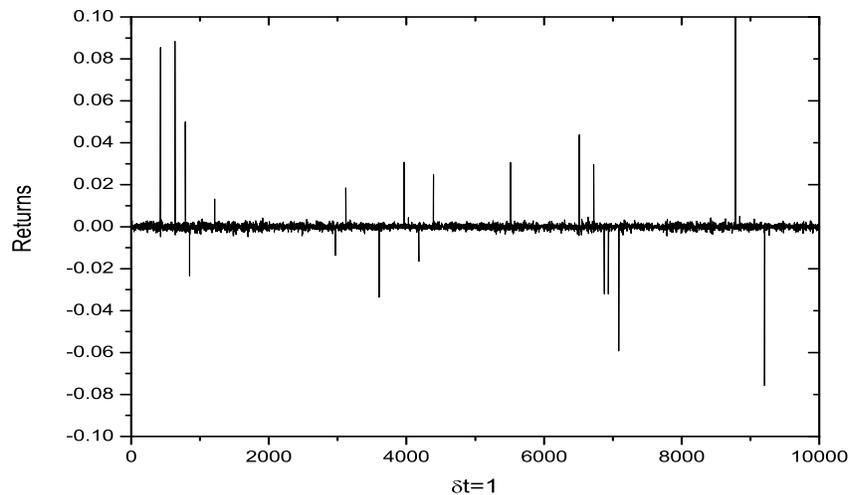}}
\caption{\label{fig:epsart} The returns of simulated price
fluctuations for $\delta t$=1. It can be seen that large events
are frequent in the fluctuations of stock market when compared
with a normal process.}
\end{figure}
\par The second, each old cluster such as \textbf{Q} collapses with the probability
 \begin{equation}
 P_{n}^{q}(t)=c^{q}(t)/L^{2}
\end{equation}
  which indicates that the probability with which a cluster collapses would increase with its
  growth. When a cluster takes up all of the sites of the
  lattice, it would surely collapse.
 Once the old cluster collapses, the members of the new clusters whose states are not necessarily the same as the old one are
 produced with fixed probability $P_{h}$ and take up the sites where the old cluster has existed.
 \par The final circumstance, clusters sleep with the probability $1-P_{d}(t)-P_{n}(t)$.
\par (4) Repeat step (2) and (3) for enough times.

\section{Simulation Results}

  The typical parameter space we adopt in our simulation is as follows:
  $l=0.0001$, $k=0.0001$, $N_{T}=50$,
 $P_{h}=0.01$, $a(0)=0.35$, $P_{d}(0)=1.0$, $N=100$, $\mu=0.59$, $\nu_{1}=0.00005$ and $\nu_{2}=0.0001$. About $100$
 traders (in other words, $P_{in}$=0.01)
are distributed on a square lattice randomly and the initial stock
price is $1.0$. The simulation results are very sensitive to some
of the parameters such as $l$, $k$, $\mu$, $\nu_{1}$ and
$\nu_{2}$. When the values of them are little larger, the price
fluctuations would be very exquisite and when they are little
smaller, the price trend would be very meek. But the simulation
results are not very sensitive to other parameters for other
values of them could lead to the results which are in good
agreement with the real data, too. But there are some amazing
results we should point out: (1) $\mu$ is not 0.5 but 0.59, little
larger than 0.5, which is very close to the threshold value: 0.593
in the percolation theory which may be a support to the point that
the real markets should operate close to the critical point where
profitable trade opportunities are barely detectable \cite{50}.
The process by which the market self-organizes close to the
critical point is more likely to be of evolutionary nature and
hence to take place on longer time scales \cite{51}. And this
result suggests that the choice whether to buy or to sell is not
completely random as the traditional point stands which implies
that $\mu$ should be 0.5. (2) From the value space we could see
that $\nu_{2}=2\nu_{1}$, which means that the affliction which is
brought by losing one unit of wealth would be twice as much as the
satisfaction which is brought by gaining the same amount of wealth
according to Kahneman's Prospect Theory \cite{52}.
\par To compare
the statistical properties of the price generated by our model and
that of real stock markets further, we study the returns of price
which is defined as Equ.(10):
\begin{equation}
Z(t)=\log P(t+\delta t)-\log P(t)
\end{equation}
\begin{figure}
\scalebox{1.2}[1.2]{\includegraphics{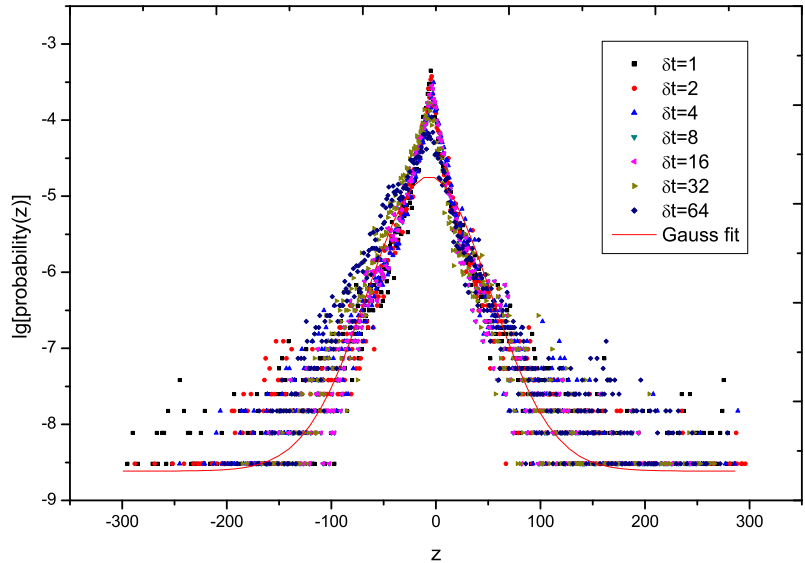}}
\caption{\label{fig:epsart}The probability distributions of price
returns with $\delta t$=1,2,4,8,16,32,64 respectively. In this
figure, the central part of the distribution of returns appears to
be well fitted by a l\'{e}vy distribution.}
\end{figure}

\begin{figure}
\scalebox{1.2}[1.0]{\includegraphics{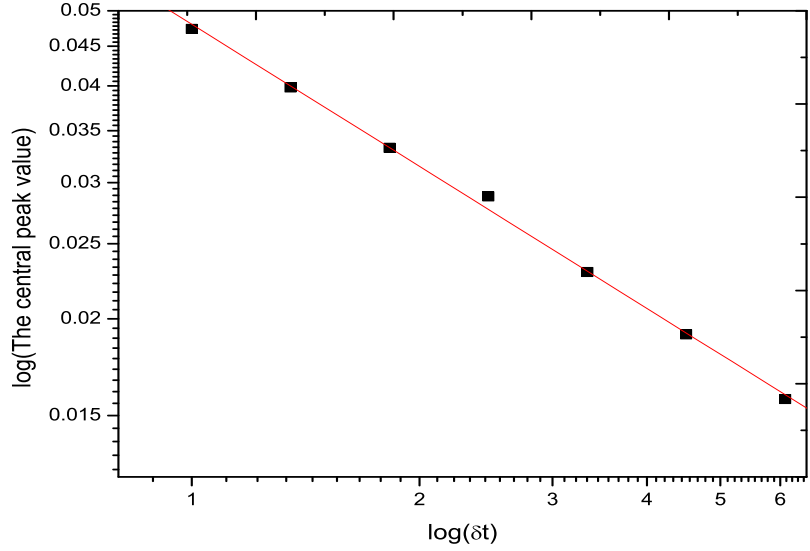}}
\caption{\label{fig:epsart}The central peak value as a function of
$\delta t$. The slope of the fitted line is $-0.61\pm0.01$ which
is very close to the real value 0.62 found in Hang Seng index
\cite{36}.}
\end{figure}

\begin{figure}
\scalebox{1.2}[1.0]{\includegraphics{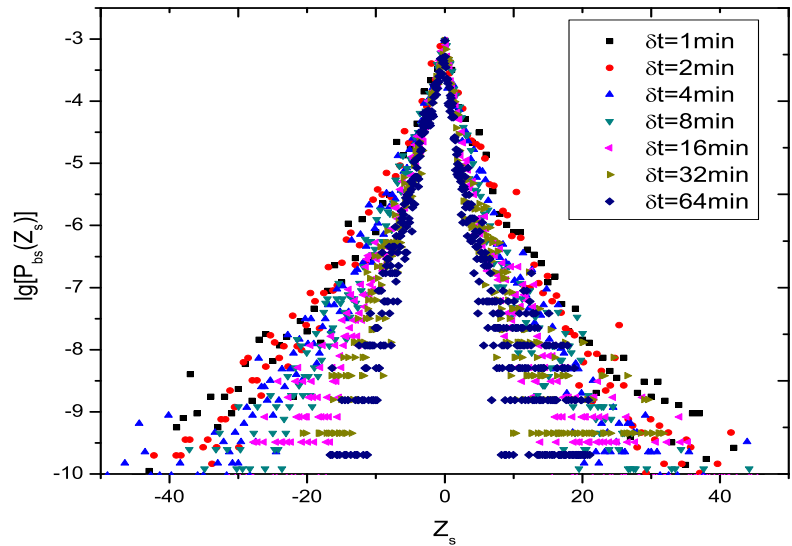}}
\caption{\label{fig:epsart}Re-scaled plot of the probability
distributions shown in figure 4. Data collapse is evident after
using re-scaled variables with $\alpha=1.61$. The abscissa is the
re-scaled returns, the ordinate is the logarithm of re-scaled
probability.}
\end{figure}
 Mandelbrot has proposed that the distribution of returns is
consistent with a L\'{e}vy stable distribution \cite{33}. In 1995,
Mantegna and Stanley analyzed a large set of data of the S\&P500
index. It has been reported that the central part of the
distribution of S\&P500 returns appears to be well fitted by a
L\'{e}vy distribution, but the asymptotic behavior of the
distribution shows faster decay than that predicted by a L\'{e}vy
distribution \cite{25,35}. The similar characteristic of the
distribution of returns is also found in Hang Seng index
\cite{36}.
   Figure 4 shows the probability density of normalized returns, which display a clear L\'evy
   distribution for $\delta t$=1, 2, 4, 8, 16, 32, 64.

\par Because larger $\delta t$ implies less data points, it is
difficult to determine the parameters characterizing the
distributions only by investigating the spreads. Hence, we studied
the peak values at the center of the  distributions, i.e., the
probability of zero return $P_{b}(Z=0)$ as the function of $\delta
t$. With this choice, we can investigate the point of each
probability distribution which is least affected by noise. Figure
5 shows the central peak value versus $\delta t$ in a
   log-log plot. It can be seen that all the data is well fitted by
   a straight line with the slope $-0.61\pm0.01$ which is very close
    to the real value 0.62 found in Hang Seng index \cite{36}. This observation agrees
   with theoretical model leading to a L\'{e}vy distribution.
\par If we assume that the central part of the distribution of
returns can be described by a L\'{e}vy stable symmetrical
distribution with an index $\alpha$ and parameter $\gamma$,
\begin{equation}
P_{b}^{\alpha}(Z,\delta
t)\equiv(1/\pi)\int_{0}^{\infty}e^{-\gamma\delta t
|q|^{\alpha}}\cos(qZ)dq
\end{equation}
where $e^{-\gamma\delta t |q|^{\alpha}}$ is the characteristic
function of a L\'{e}vy symmetrical stable process, the probability
of zero return is given by
\begin{equation}
P_{b}(0)=P_{b}^{\alpha}(0,\delta
t)=\Gamma(1/\alpha)/[\pi\alpha(\gamma\delta t)^{1/\alpha}]
\end{equation}
where $\Gamma$ is the Gamma function. Using the value
$-0.61\pm0.01$ for the slope of the fitted line to the data
(figure 5), we obtain the index $\alpha = 1.61\pm0.02$.
\par To check whether the L\'{e}vy scaling can be extended to the
entire probability distribution of returns generated by our model,
we notice that under the transformation:
\begin{equation}
Z_{s}\equiv Z/[(\delta t)^{(1/\alpha)}]
\end{equation}
and
\begin{equation}
P_{bs}(Z_{s})\equiv(\delta t)^{1/\alpha}P_{b}^{\alpha}(Z,\delta
t)=(\delta t)^{1/\alpha}P_{b}^{\alpha}[(\delta
t)^{1/\alpha}Z_{s},\delta t]
\end{equation}
the distributions for different time scales $\delta t$ will
collapse onto one curve. Figure 6 shows the re-scaled
distributions for the same data in figure 4 in the scaled
variables, i.e., $P_{bs}(Z_{s})$ versus $Z_{s}$. Data collapse is
evident, except for some data points in the tails for large
$\delta t$. The closer to the central point $Z_{s}=0$, the
stronger is the extent of data collapse. These observations imply
that the L\'{e}vy distribution is a better description of the
dynamics of the random process underlying the variation of returns
in the central part of the probability distribution $P_{b}(Z)$
over $\delta t$ spanning at least two orders of magnitude.

\begin{figure}
\scalebox{1.2}[1.0]{\includegraphics{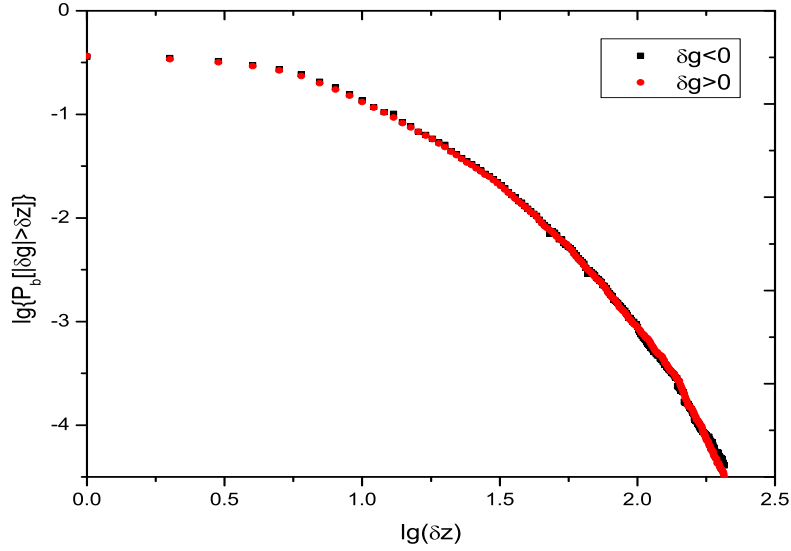}}
\caption{\label{fig:epsart}The accumulate probability
distributions $P_{b}(g>Z)$ of 1-minute returns generated by our
model. For data in the region $10\leq Z \leq 200$, regression fits
yield $\alpha=2.93$ (positive tail) and $\alpha=2.78$ (negative
tail).}
\end{figure}

\par In order to determine if an exponential truncated L\'{e}vy
distribution can be used to describe the stochastic process and to
investigate the kind of asymptotic behavior outside the L\'{e}vy
stable region, we study the accumulate distribution $P_{b}(g>Z)$
of the fluctuations of financial data.
\par For a stable symmetric L\'{e}vy distribution ($0<\alpha<2$),
the two tails show a power-law asymptotic behavior
\begin{equation}
P_{b}(Z)\thicksim Z^{-(1+\alpha)}
\end{equation}
and hence the second moment diverges. This leads to an asymptotic
power-law for the accumulate distribution for both the positive
and negative tails \cite{33} in the form
\begin{equation}
P_{b}(g>Z)\thicksim Z^{-\alpha}
\end{equation}
\par Figure 7 shows the accumulate probability distribution of
returns $P_{b}(g>Z)$ for $\delta t=1$ min. for the data generated
by our model. For data in the region $10\leq Z \leq 200$,
regression fits yield $\alpha=2.93$ (positive tail) and
$\alpha=2.78$ (negative tail). These results appear to be outside
the L\'{e}vy stable range of $0<\alpha<2$ but they fit well the
result produced from the real data which is near 3.

\section{Conclusion}
Compared with the Cond-Bouchaud percolation model, our model
presents a nonconservation  market. With the evolution of the
model's topology, there are new traders coming in and old ones
leaving, which depicts the real stock markets more approximately.
What is more, the process of the amalgamation and expansion, and
the breakdown of the clusters in our model well consists with the
phenomena (so called herd behaviors) in the real stock market that
more and more people would take the same performance when they
found more and more people around them take the same action. Some
other simulations show that the information entropy, when we
consider the clusters as various information resources and the
process of merging and collapsing as the spreading and dying out
of the information, has some relationships with the point which we
define as the break-point of price. Moreover, there are also some
amazing facts in the results of our simulations such as why the
first term on the right hand of Equ.(4) approximate to the
threshold value in the percolation theory. Since the main goal of
this article is to establish and describe the model itself, we
would not give detailed simulation results and analysis, which
will be given elsewhere soon.

\section{Acknowledgement}
This work has been partially supported by the National Natural
Science Foundation of China under Grant No.70171053 and
No.70271070, the Specialized Research Fund for the Doctoral
Program of Higher Education (SRFDP No.20020358009), and the
Foundation for graduate students of University of Science and
Technology of China under Grant No. USTC-SS-0501.

\end{document}